\documentclass[aps,prb,twocolumn,showpacs,superscriptaddress,groupedaddress]{revtex4}  
\usepackage{graphicx}  
\usepackage{dcolumn}   
\usepackage{bm}        
\usepackage{amssymb}   
\usepackage{color}
\usepackage[colorlinks = true,
            linkcolor = blue,
            urlcolor  = blue,
            citecolor = blue,
            anchorcolor = blue]{hyperref}
\usepackage[all]{hypcap}

\usepackage[applemac]{inputenc}
\usepackage{amsmath}
\usepackage{mathptmx}

\usepackage[section,subsection,subsubsection]{extraplaceins}

\newcommand{\sdo}{SrDy$_2$O$_4$}
\newcommand{\sho}{SrHo$_2$O$_4$}
\newcommand{\seo}{SrEr$_2$O$_4$}
\newcommand{\mub}{$\mu_\text{B}$ }
\newcommand{\uud}{$\uparrow \uparrow \downarrow$}

\newcommand{\uuddud}{$\uparrow \uparrow \downarrow \downarrow \uparrow \downarrow$}
\newcommand{\refsub}[2][{}]{\hyperref[#2]{\ref{#2}(#1)}} 

\begin{document}

%



\title{Field dependence of the magnetic correlations of the frustrated magnet \sdo}
\author{N. Gauthier}  
\email{nicolas.gauthier4@gmail.com}
\affiliation{Laboratory for Scientific Developments and Novel Materials, Paul Scherrer Institut, 5232 Villigen, Switzerland}

\author{A. Fennell} 
\affiliation{Laboratory for Neutron Scattering and Imaging, Paul Scherrer Institut, 5232 Villigen, Switzerland}

\author{B. Pr\'evost}
\affiliation{D\'epartement de physique, Universit\'e de Montr\'eal, Montr\'eal, Canada}\affiliation{Regroupement Qu\'eb\'ecois sur les Mat\'eriaux de Pointe (RQMP)}

\author{A. D\'esilets-Benoit}
\affiliation{D\'epartement de physique, Universit\'e de Montr\'eal, Montr\'eal, Canada}\affiliation{Regroupement Qu\'eb\'ecois sur les Mat\'eriaux de Pointe (RQMP)}

\author{H.A. Dabkowska} 
\affiliation{Brockhouse Institute for Materials Research, Hamilton, Canada}

\author{O. Zaharko}
\affiliation{Laboratory for Neutron Scattering and Imaging, Paul Scherrer Institut, 5232 Villigen, Switzerland}

\author{M. Frontzek}
\affiliation{Laboratory for Neutron Scattering and Imaging, Paul Scherrer Institut, 5232 Villigen, Switzerland}
\affiliation{Quantum Condensed Matter Division, Oak Ridge National Laboratory, 37831 Oak Ridge, TN, USA}

\author{R. Sibille}
\affiliation{Laboratory for Neutron Scattering and Imaging, Paul Scherrer Institut, 5232 Villigen, Switzerland}

\author{A.D. Bianchi} 
\affiliation{D\'epartement de physique, Universit\'e de Montr\'eal, Montr\'eal, Canada}\affiliation{Regroupement Qu\'eb\'ecois sur les Mat\'eriaux de Pointe (RQMP)}

\author{M. Kenzelmann}
\email{michel.kenzelmann@psi.ch}
\affiliation{Laboratory for Scientific Developments and Novel Materials, Paul Scherrer Institut, 5232 Villigen, Switzerland}
\vskip 0.25cm
\date{\today}

\begin{abstract}

The frustrated magnet \sdo\ exhibits a field-induced phase with a magnetization plateau at $1/3$ of the saturation value for magnetic fields applied along the $b$-axis. We report here a neutron scattering study of the nature and symmetry of the magnetic order in this field-induced phase. Below $T\approx 0.5$~K, there are strong hysteretic effects, and the order is short or long ranged for zero-field and field cooling, respectively. We find that the long-range ordered magnetic structure within the zig-zag chains is identical to that expected for the one-dimensional axial next-nearest neighbour Ising (ANNNI) model in longitudinal fields. The long-range ordered structure in field contrasts with the short-range order found at zero field, and is probably reached through enhanced quantum fluctuations with increasing fields.


\end{abstract}

\maketitle




\section{Introduction}
In frustrated magnets with competing interactions, phases with different magnetic order can have a similar energy. As a result, small perturbations can strongly affect the magnetic properties, and strong hysteretic effects can occur. In particular, an applied magnetic field can act as a tuning parameter between different states. Therefore, frustrated magnets often exhibit complex field-dependent phase diagrams. 
The field-induced phases can reveal themselves as plateaux in the field-dependent magnetization and they give rise to strong correlations and unusual properties. For example in the prototypical spin ice Dy$_2$Ti$_2$O$_7$, a magnetization plateau is observed for fields applied along $(111)$. The ground state related to this plateau is called Kagom\'e ice and has a macroscopic degeneracy.\cite{Matsuhira2002,Gardner2010}

%
%
%
%
%
%

Magnetization plateaux have been observed in \sdo, \sho\ and \seo, which are part of the Sr\textit{R}$_2$O$_4$ (\textit{R}~=~rare earth) family of frustrated magnets.\cite{J.Hayes2012} In many members of this family, the magnetism is dominated by magnetic ions forming zig-zag chains along the $c$-axis.\cite{Hayes2011,Quintero-Castro2012,Fennell2014,Wen2015} There are two crystallographically inequivalent zig-zag chains present, and these have different moment anisotropies.\cite{Petrenko2014,Fennell2014,Malkin2015} The zig-zag chains can be modelled by the one-dimensional (1D) axial next-nearest-neighbour Ising (ANNNI) model with exchange interactions $J_1$ and $J_2$, corresponding to the nearest and next-nearest neighbour interactions, respectively.\cite{Selke1988} For the case where both $J_1$ and $J_2$ are antiferromagnetic, this model predicts the existence of a magnetization plateau in longitudinal field at $\frac{1}{3}$ of the saturation value $M_s$ that corresponds to an ordering \uud\ along the chain.\cite{Oguchi1965,Morita1972,Rujan1983} 

In \sdo, a plateau at $\frac{1}{3}M_s$ is observed for fields applied along the $b$-axis suggesting the existence of the \uud\ state. At $T~=~0.5$~K, this plateau extends from $H_{c1}~=~0.16$~T to $H_{c2}~=~2.03$~T. \cite{J.Hayes2012} Specific heat and ultrasound measurements confirm a field-induced phase between these critical fields.\cite{Cheffings2013,Bidaud2016} This field-induced order contrasts with absence of long range order down to $T~=~20~$mK in zero field.\cite{Fennell2014,Gauthier2017a} The presence of only short range correlations at the lowest temperatures in zero field has been justified by the trapping of slowly decaying defects due to a dimensionality crossover.\cite{Gauthier2017} We report here a detailed characterization of the field-induced order by single crystal neutron diffraction. Our measurements reveal the presence of strong hysteretic effects and the existence of a three-dimensional (3D) long range order with a \uud\ spin order along the chains. At fields higher than the plateau phase, a partial ferromagnetic state exists where the chains in longitudinal fields are fully polarized.

\section{Experimental details}
\sdo\ single crystals were prepared as described previously.\cite{Balakrishnan2009,Gauthier2017} Magnetic order as function of magnetic field was measured on the TriCS single crystal neutron diffractometer, SINQ at the Paul Scherrer Institut. The sample used for the determination of the magnetic structure had an ellipsoid shape in order to reduce the anisotropy of neutron absorption, while the remaining samples had the shape of a rectangular prism.
The samples were aligned in the $ac$ plane and inserted into a dilution refrigerator in a 6~T vertical magnet, with the field along the $b$-axis. For the determination of the field-induced magnetic structure on TriCS, datasets were collected using two different neutron wavelengths, $\lambda~=~2.317~\rm{\AA}$ which is strongly affected by neutron absorption due to Dy, and $\lambda~=~1.178~\rm{\AA}$ with smaller absorption effects. For $\lambda~=~1.178~\rm{\AA}$, the limited out-of-plane resolution lead to peak overlap along the $k$ direction. Scans were measured along $k$ to evaluate this overlap and correct for it. Both datasets were corrected for absorption, which was estimated through a finite element analysis based on the sample geometry. The same conclusions are reached from the datasets of both neutron wavelengths. The field-induced phase diagram was also investigated using the DMC neutron diffractometer at SINQ. Several single crystals were coaligned in the $\textit{bc}$ plane and fixed with Araldite glue on a silicon plate mounted on a copper holder. The sample was inserted in a dilution refrigerator in a 2~T horizontal magnet, with the field along the $b$-axis. Measurements were carried out using a neutron wavelength $\lambda~=~3.8$~\AA.

DC susceptibility measurements were done in a Quantum Design MPMS equipped with an iHelium $^3$He option.
The sample with dimensions $0.9 \times 1.9 \times 0.4$~mm$^3$ was measured at $T~=~0.66$~K with the field applied along the $b$-axis. The measurements were corrected for demagnetization with the demagnetization factor $N~=~0.128$, calculated from the sample dimensions using the equations for a rectangular prism.\cite{Aharoni1998}


\section{Experimental results}

\subsection{Field-induced magnetic order}
\label{magstruc}

\begin{figure}[!htb]
\includegraphics[scale=0.8]{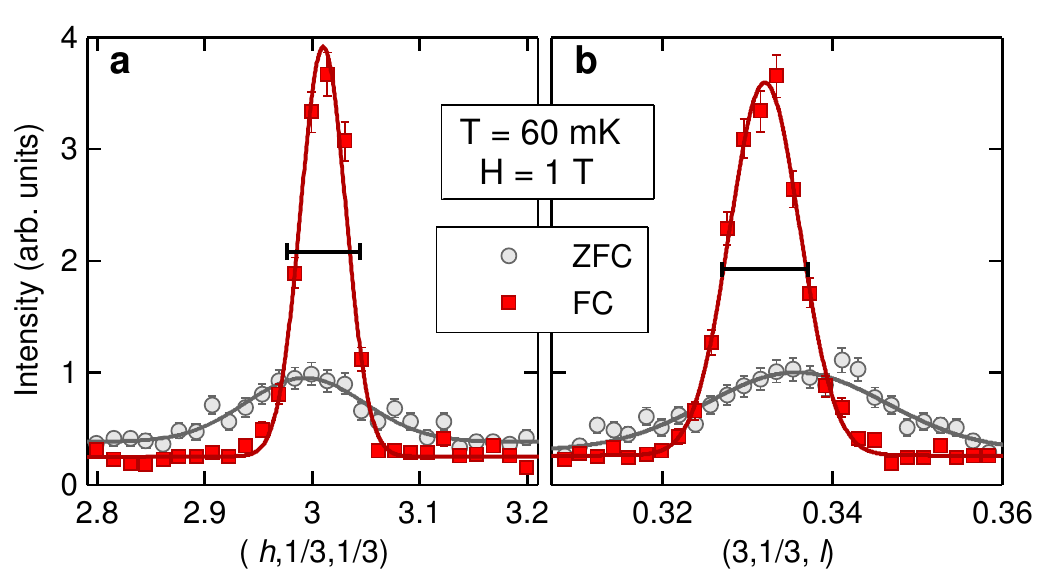}
\caption{ (a) $h$-scan and (b) $l$-scan of the magnetic scattering near $\textbf{Q}~=~(3,\frac{1}{3},\frac{1}{3})$ at $T~=~60$~mK and $H~=~1$~T in zero-field cooled (ZFC) and field cooled (FC) conditions. The horizontal black line corresponds to the instrumental resolution estimated from the nuclear peak $\textbf{Q}~=~(4,0,0)$ at $T~=~2$~K and $H~=~0$~T.}
\label{TRICS/peaks}
\end{figure}

The magnetic field-induced order corresponding to the $\frac{1}{3}M_s$ magnetization plateau for fields along the $b$-axis in \sdo\ was investigated by single crystal neutron diffraction measurements. The previous studies of this field-induced order, by specific heat and ultrasound measurements,\cite{Cheffings2013,Bidaud2016} could not establish whether the order has short or long range correlations. Neutron scattering can quantify the correlation length from the widths of the Bragg peaks and we show here that the system reaches a short range order in zero-field cooling (ZFC) conditions and a long range order in field-cooled (FC) conditions. Scans along $h$ and $l$ of the magnetic Bragg peak $\textbf{Q}~=~(3,\frac{1}{3},\frac{1}{3})$ illustrate narrow peaks for FC conditions and broad ones for ZFC conditions (Fig.~\ref{TRICS/peaks}). The widths of the magnetic Bragg peaks in FC conditions are similar to the instrumental resolution estimated from the nuclear Bragg peak $\textbf{Q}~=~(4,0,0)$ measured out of the field-induced order ($T~=~2$~K and $H~=~0$~T). This indicates that they are resolution limited and that FC conditions establishes a long range field-induced order. 

A phase diagram of the field-induced order was created from the neutron scattering amplitude at $\textbf{Q}~=~(0,0.66,0.67)$, as shown in Fig.~\refsub[a]{TriCS/allData_log}. The data collection was performed with the TriCS diffractometer by increasing the field $H$ at various temperatures $T$ starting from ZFC conditions. The boundaries of this phase, indicated by the lower critical field H$_{c1}$ and the upper critical field $H_{c2}$, are in good agreement with specific heat and ultrasound measurements.\cite{Cheffings2013,Bidaud2016} Strong hysteretic effects appear below $T\approx0.5$~K and will be described in details in section~\ref{hysteresis}. 

\begin{figure}[!htb]
\includegraphics{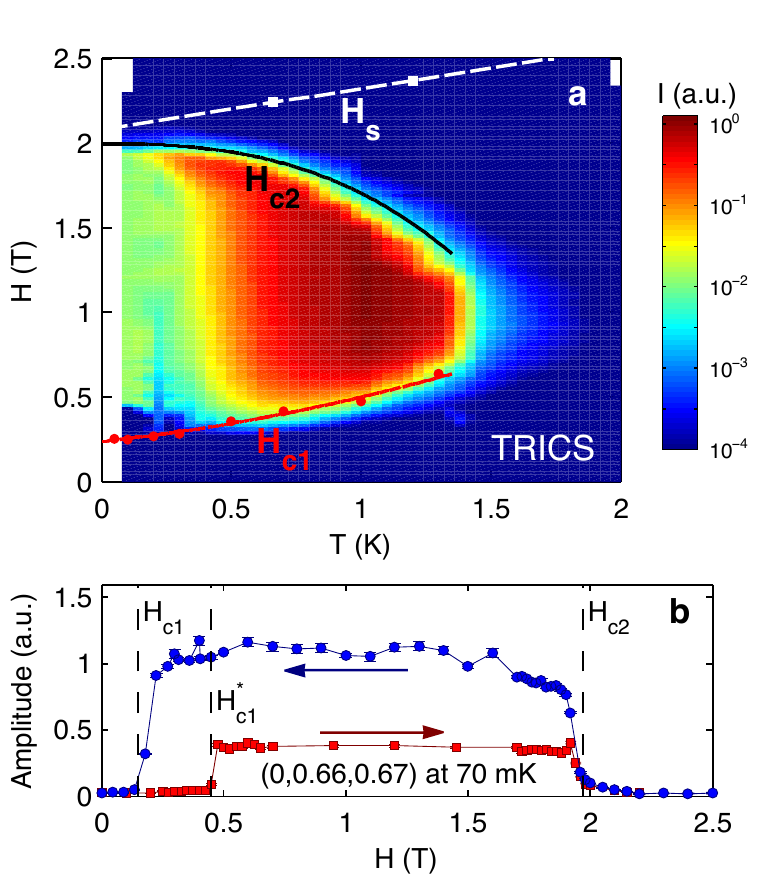}
\caption{(a) $HT$-phase diagram for $\textbf{H} \parallel \textbf{b}$ of the field induced phase compiled from the peak intensity of the magnetic peak $\textbf{Q}~=~(0,0.66,0.67)$ after zero-field cooling with the intensity shown on a logarithmic scale. All data were collected at fixed temperature with increasing magnetic fields on TriCS. The red circles show $H_{c1}$ for decreasing fields in FC conditions measured on DMC. The white squares show the saturation field $H_s$ obtained from magnetization measurements without correction for demagnetization effects. These demagnetization effects are comparable for the experiment with TriCS due to similar sample shapes. (b) Magnetic field hysteresis at 70~mK of the amplitude of the magnetic peak $\textbf{Q}~=~(0,0.66,0.67)$.}
\label{TriCS/allData_log}
\end{figure}

\begin{figure}[!htb]
\includegraphics[scale=1.1]{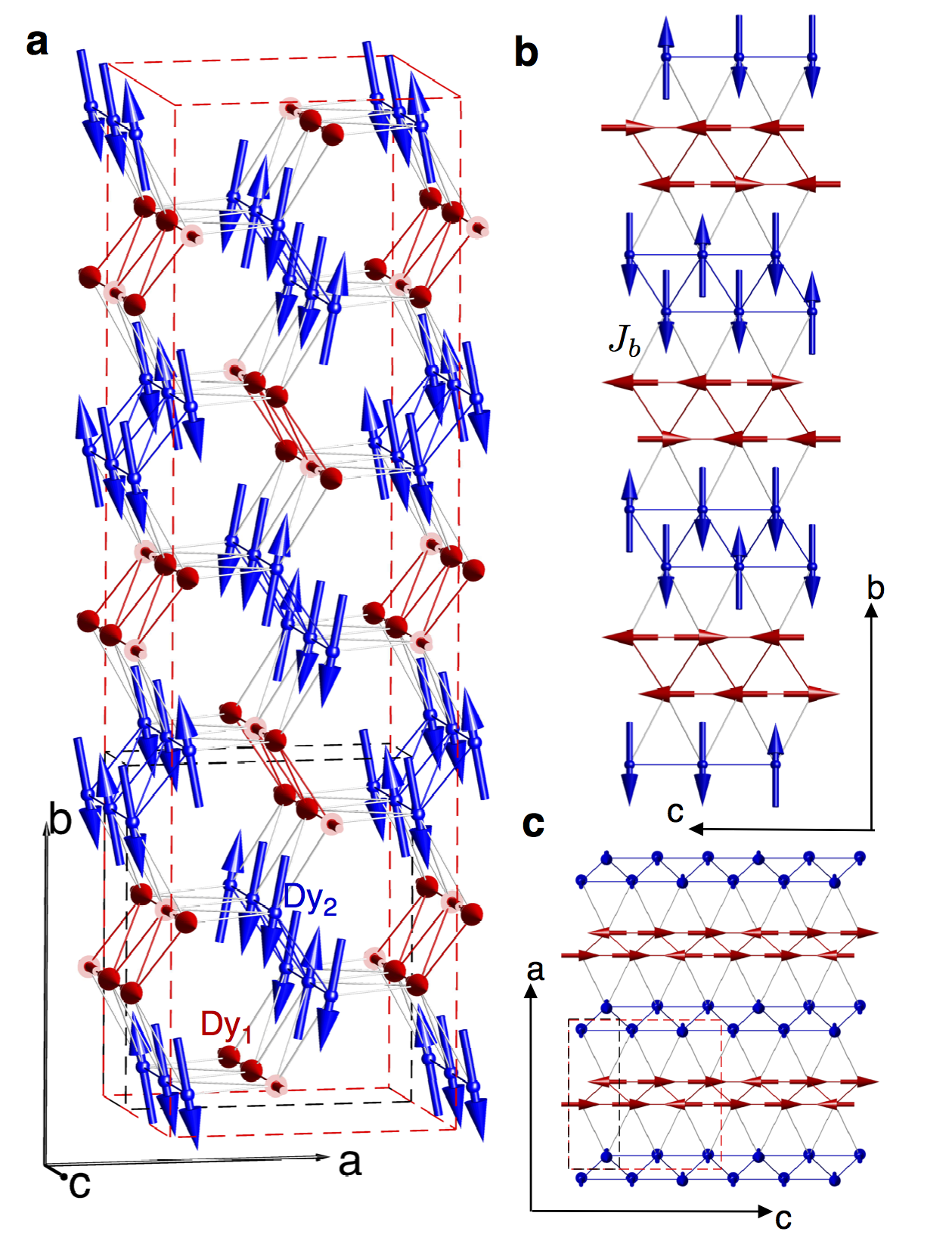}
\vspace{-0.7cm}
\caption{Magnetic structure of \sdo\ in the field-induced phase. The $\textbf{k}_0$ and $\textbf{k}_{1/3}$ propagation vectors are used to describe this structure. The chemical unit cell is outlined in gray lines and the magnetic unit cell in red lines. The inequivalent sites 1 and 2 are illustrated in red and blue, respectively.}
\label{DMC/MagStruct}
\end{figure}

The field-induced long range order has two types of magnetic scattering: a ferromagnetic contribution at the nuclear Bragg peak positions, described by a propagation vector $\textbf{k}_0~=~(0,0,0)$ and an antiferromagnetic contribution described by a commensurate propagation vector $\textbf{k}_{1/3}~=~(0,\frac{1}{3},\frac{1}{3})$. The magnetic structure is described by a multi-\textbf{k} structure with both propagation vectors $\textbf{k}_0$ and $\textbf{k}_{1/3}$, and is illustrated in Fig.~\ref{DMC/MagStruct}. The dataset for this structure determination was collected at  $T~=~0.1$~K and $H~=~1.0$~T in FC conditions using $\lambda~=~1.178~\rm{\AA}$. In FC conditions, the $\textbf{k}_{1/3}$ propagation vector  is temperature independent, and the magnetic Bragg peak amplitudes are field independent at $T~=~0.7$~K. This indicates that there is no change of the magnetic structure within the long range ordered phase.


The magnetic structure presented in Fig.~\ref{DMC/MagStruct} was determined using two basic assumptions. The first assumption is that the moments are aligned along the easy axes determined previously.\cite{Gauthier2017} There are two crystallographically inequivalent magnetic ion sites in the structure of \sdo. We refer to them as site 1 and site 2 and they are shown in Fig.~\ref{DMC/MagStruct} in red and blue, respectively. Our previous work has shown that site 2 has an easy-axis near the $b$-axis and site 1 has a weak $XY$ anisotropy in the $ac$ plane, with a favorable alignment along the $c$-axis.\cite{Gauthier2017} For the following, the moment on site 1 is considered to point along the $c$-axis and a component along the $a$-axis is discussed later. 
The second assumption used for the magnetic structure determination is that all the moments on site 1 have the same size, as do all the moments on site 2. The first assumption implies that the structure defined by the propagation vector $\textbf{k}_{1/3}$ is amplitude modulated. The second assumption requires the use of both propagation vectors and restricts the phase factor of the $\textbf{k}_{1/3}$ structure such that for every two sites with magnetic moments of $-M/2$ there is one site with magnetization $M$. Adding a ferromagnetic component of $-M/4$ from the $\textbf{k}_0$ structure then generates equal moment sizes of $3M/4$ for all sites. With these restrictions, the parameters of the $\textbf{k}_{1/3}$ and $\textbf{k}_0$ structures are highly dependent. 

The scattering corresponding to $\textbf{k}_0$ combines magnetic and nuclear contributions, so the accuracy of the refinement of the ferromagnetic component is limited. The related parameters were therefore determined from the refinement of the $\textbf{k}_{1/3}$ structure. A crystal structure refinement was performed using nuclear Bragg peaks collected at $T~=~2$~K and $H~=~0$~T ($\chi^2~=~9.63$) and the obtained scaling factor was used to determine absolute moment sizes. The presence of two domains ($\textbf{k}_{1/3}$ and $\textbf{k'}_{1/3}~=~(0,\frac{1}{3},-\frac{1}{3})$) with unequal populations was also taken into account to obtain the correct scaling between the nuclear and magnetic phases.

\begin{table}[!htb]
\begin{tabular}{c|cc}
  \hline      
     ~~~~~~~~~~~~~~                 &  ~~~~~~~~~~~~~~$\Gamma_1$ ~~~~~~~~~~~~~~       & ~~~~~~~~~~~~~~$\Gamma_2$~~~~~~~~~~~~~~       \\
\hline
 $(x,y,z)$ & $(M_x,M_y,M_z)$ & $(M_x,M_y,M_z)$ \\
 $(-x+\frac{1}{2},y-\frac{1}{2},z+\frac{1}{2})$ & $(M_x,-M_y,-M_z)$ & $(-M_x,M_y,M_z)$ \\ 
    \hline      
    \end{tabular}
\caption{The three basis vectors represented as $M_{x,y,z}$ are listed for the two 1D irreducible representations with the propagation vector $\textbf{k}_{1/3}~=~(0,\frac{1}{3},\frac{1}{3})$ at Wyckoff position $4c$ in space group \it{Pnam}.}
\label{table:irreps}
\end{table}

\begin{figure}[!htb]
\includegraphics[scale=1.1]{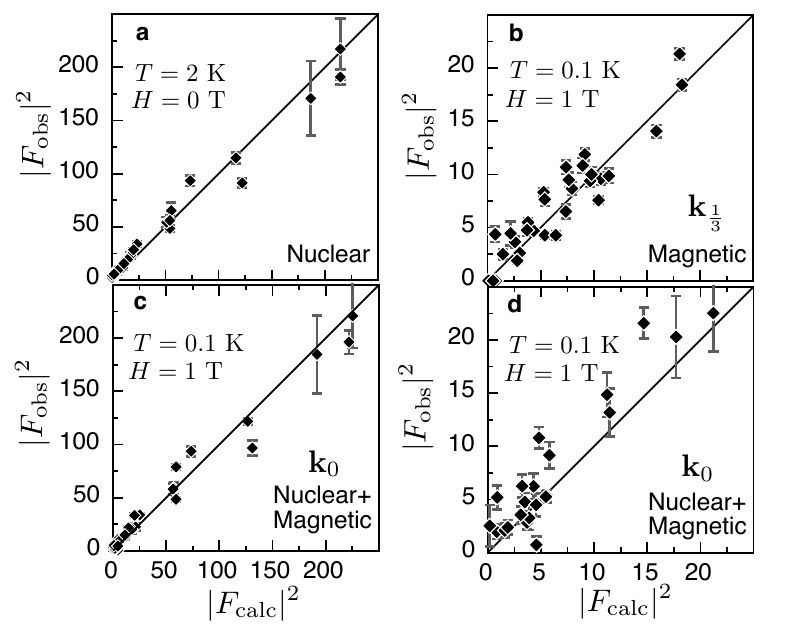}
\vspace{-0.5cm}
\caption{(a) Result of the nuclear structure refinement at $T~=~2$~K and $H~=~0$~T, represented as the observed squared structure factors $|F_\text{obs}|^2$ versus the calculated squared structure factors $|F_\text{calc}|^2$. (b)-(d) Result of the magnetic structure refinement at $T~=~0.1$~K and $H~=~1$~T for both propagation vectors $\textbf{k}_0$ and $\textbf{k}_{1/3}$. Graphs on the left show the full dataset and graphs on the right focus to the small $|F_\text{obs}|^2$ values.
}
\label{TRICS/ref}
\end{figure}

The representation analysis for the propagation vector $\textbf{k}_{1/3}$ using \textsc{BasIreps} \cite{Fullprof} shows that each inequivalent site splits into two orbits with two possible 1D irreducible representations, presented in Table~\ref{table:irreps}. While satisfying the two assumptions presented above, only the representation $\Gamma_2$ can generate a non-zero magnetic moment along the $b$-axis, as required by the magnetization measurements. The best agreement with the measurements is obtained with the \uud\ state on both inequivalent sites with moments pointing along their respective easy axis. The detailed structure is shown in Fig.~\ref{DMC/MagStruct}. The refined magnetic moments are 3.3(7)~\mub along the $c$-axis for site 1 illustrated in red and 8.4(6)~\mub nearly along the $b$-axis for site 2 illustrated in blue. The corresponding $\chi^2$ values $\chi^2({\textbf{k}_{1/3}})~=~14.04$ and $\chi^2({\textbf{k}_0)}~=~10.54$ are of the same order as the value $\chi^2~=~9.63$ obtained by refining the nuclear structure. A comparison of the calculated and observed structure factor for these three refinements is presented on Fig.~\ref{TRICS/ref}. A refinement using the representation $\Gamma_1$ instead of $\Gamma_2$ leads to a worse agreement with the observed intensities.  

Even though a weak $XY$ anisotropy in the $ac$ plane was suggested for site 1, magnetic correlations of moments along the $a$-axis were not observed experimentally.\cite{Gauthier2017} This is also the case for the field-induced magnetic structure presented here: the presence of a component along the $a$-axis does not improve significantly the $\chi^2$ value and in that case its size is negligible compared to the component along the $c$-axis. 

The magnetic order on site 1 for a single domain leads to a finite magnetization along the $c$-axis for magnetic fields applied along the $b$-axis. This net magnetization component originating from site 1 is surprisingly transverse to the applied field. The second domain described by $\textbf{k'}_{1/3}$ creates a transverse magnetization in the opposite direction. In our measurements, these domains appeared in a 1:2 ratio. Due to this unequal domain population, that should create a net magnetization transverse to the direction of the applied magnetic field. Transverse magnetization measurements are needed to directly confirm this prediction. The next best solution with a larger $\chi^2({\textbf{k}_{1/3}})~=~21.25$ features magnetic order on the site 1 chain with a \uuddud\ pattern without finite transverse magnetization and with the \uud\ arrangement on site~2. 

\subsection{Hysteresis of the field-induced phase}

\label{hysteresis}

\begin{figure}[!htb]
\includegraphics[scale=0.45]{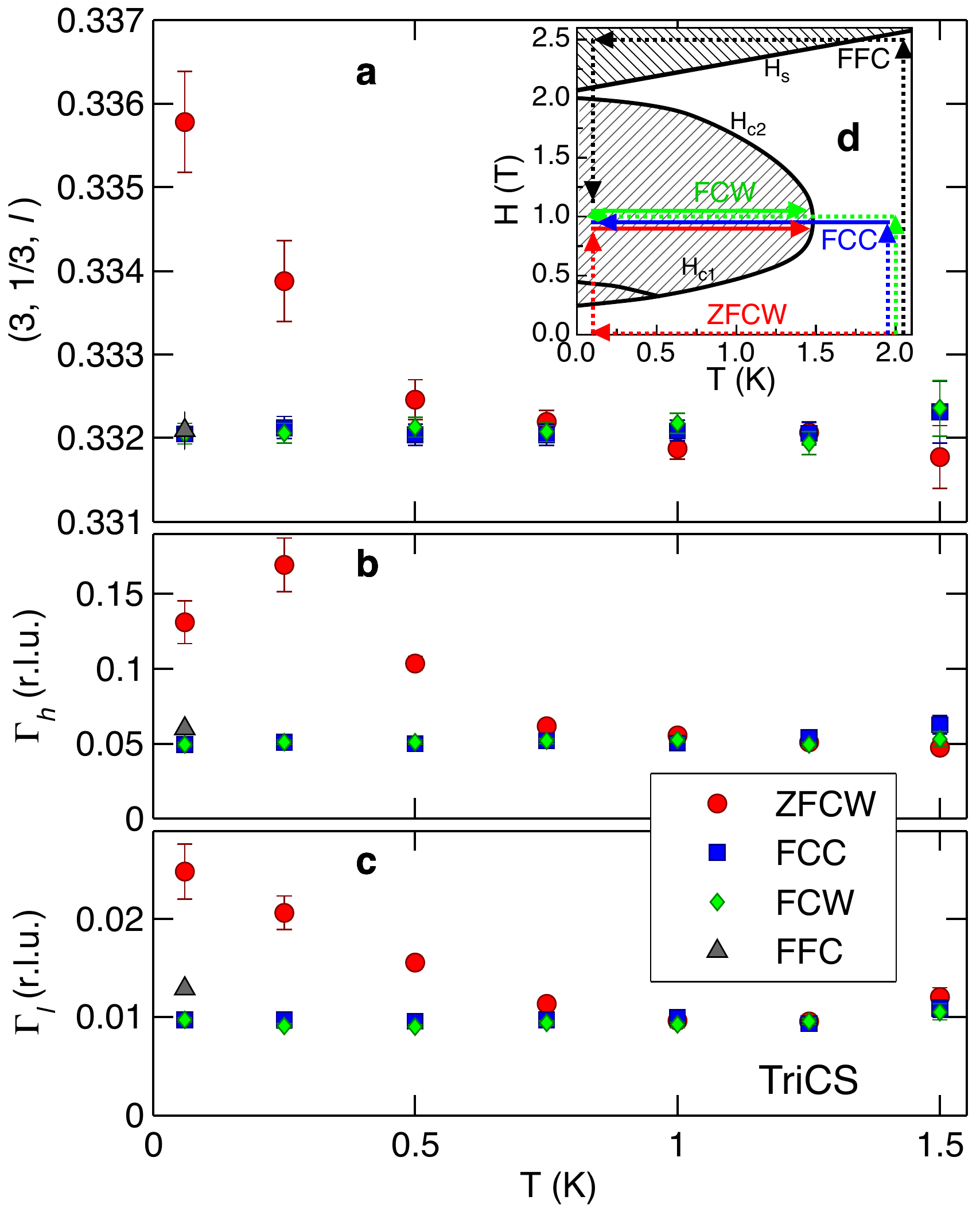}
\vspace{-0.2cm}
\caption{Temperature dependence of (a) $\textbf{Q}~=~(3,\frac{1}{3},\frac{1}{3})$ peak position and (b)-(c) width $\Gamma$ for different measurement histories: zero-field cooled on warming (ZFCW), field cooled on cooling (FCC), field cooled on warming (FCW) and field-cooled in ferromagnetic phase (FFC). The small deviation from the exact value $l~=~\frac{1}{3}$ in FCC is attributed to the precision of the sample alignment.
(d)~Detailed history trajectories in the phase diagram.}
\label{TriCS/TriCSTemp}
\end{figure}

The magnetic correlations show very strong hysteretic effect, which we studied using the magnetic Bragg peak $\textbf{Q}~=~(3,\frac{1}{3},\frac{1}{3})$ measured at $H~=~1$~T with four different histories (Fig.~\ref{TriCS/TriCSTemp}): (1) zero-field cooled on warming (ZFCW), (2) field cooled on cooling (FCC), (3) field cooled on warming (FCW) and (4) field-cooled in the ferromagnetic phase (FFC). The detailed trajectories in the phase diagram are shown in Fig.~\refsub[d]{TriCS/TriCSTemp}. The positions $h$ and $l$ and the full-widths at half maximum (FWHM) $\Gamma_h$ and $\Gamma_l$ were determined by fitting a Gaussian function to the scattering along $h$ and $l$, respectively. The correlation length is defined by $\xi~=~2/\sqrt{\Gamma_\text{exp}^2-\Gamma_\text{res}^2}$ where $\Gamma_\text{exp}$ is the FWHM of the measured scattering and $\Gamma_\text{res}$ is the FWHM expected from the instrumental resolution. 

The FCC and FCW conditions result in the longest correlation lengths and the peaks are resolution limited at $T~=~60$~mK, indicating correlation lengths along the $a$-axis $\xi_a>65~\rm{\AA}$ and along the $c$-axis $\xi_c>112~\rm{\AA}$, as estimated from $\xi>2/\Gamma_\text{res}$. In ZFCW conditions, the correlation lengths below $T~=~0.7$~K are significantly shorter, with $\xi_a~=~27(4)~\rm{\AA}$ and $\xi_c~=~48(7)~\rm{\AA}$ at $T~=~60$~mK [Fig.~\refsub[b]{TriCS/TriCSTemp}-\refsub[c]{TriCS/TriCSTemp}] and the position of the scattering maximum deviates from a commensurate value [Fig.~\refsub[a]{TriCS/TriCSTemp}]. The significantly reduced peak intensity below $T\approx0.5$~K in the phase diagram of Fig.~\refsub[a]{TriCS/allData_log} is a consequence of larger peak widths due to the short range correlations. When the field-induced state is entered from the ferromagnetic state at $T~=~0.1$~K (FFC), the correlation lengths are longer than for ZFC but still shorter than for FCC conditions. 

The temperature and field history also affects the phase boundaries at the lowest temperatures. At $T~=~70$~mK, the upper boundary is $H_{c2}~=~1.97(3)$~T and independent of the field history. The lower boundary is $H_{c1}^*~=~0.45(1)$~T for an increasing field and $H_{c1}~=~0.15(2)$~T for a decreasing field, as shown in Fig.~\refsub[b]{TriCS/allData_log}. The lower boundary obtained for decreasing fields is illustrated on the phase diagram of Fig.~\refsub[a]{TriCS/allData_log} by red circles. 

\begin{figure}[!htb]
\includegraphics[scale=1.15]{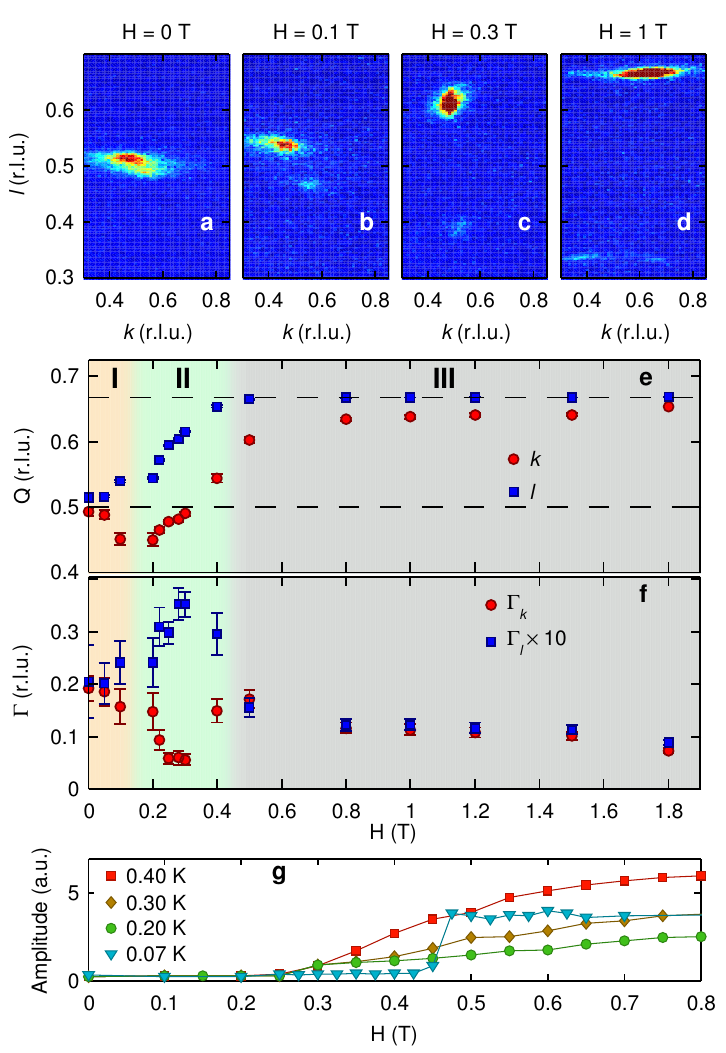}
\vspace{-0.2cm}
\caption{Field dependence at $T~=~60$~mK of (a)-(d) the diffuse scattering map in the $kl$ reciprocal plane, (e) the maximum position in $k$ and $l$ and (f) the peak width $\Gamma$ along both directions measured with DMC using $\lambda~=~3.8~\rm{\AA}$ in ZFC conditions (see text). The background colours differentiate the three regimes discussed in the text. (g) Magnetic field dependence of the scattering intensity at fixed $\textbf{Q}~=~(0,0.66,0.67)$ for different temperatures in ZFC measured with TriCS.}
\label{DMC/allDMC}
\end{figure}

Neutron scattering maps of the $kl$ reciprocal plane have been measured at $T~=~60$~mK after a zero-field cooling 
on the DMC neutron diffractometer and are shown on Fig.~\ref{DMC/allDMC} for different fields up to a maximum field $H~=~1.8$~T, smaller than $H_{c2}$. In these conditions, measurements taken at $H~=~0.3$~T approaching from smaller or larger fields are indistinguishable and indicate that the system remained in the short range correlations regime throughout the measurements.
The diffuse scattering near $\textbf{Q}~=~(0, 0.5, 0.5)$ in zero field shifts in $k$ and $l$ in finite field and splits in two peaks reaching $\textbf{k}_{1/3}$ at $H~=~1$~T. This gradual evolution from the zero field short range order to the field-induced order suggests a gradual reordering of magnetic moments, possibly by individual spin flips. 


The peak positions $\textbf{Q}$ and widths $\Gamma$ as function of applied magnetic field were obtained from the $kl$ reciprocal plane maps with $h~=~0$: the scattering has been integrated along one direction ($k$ or $l$) and fitted by a Lorentzian function along the other direction. The fitted positions $k$ and $l$ and widths $\Gamma_k$ and $\Gamma_l$ are illustrated in Fig.~\refsub[e]{DMC/allDMC}-\refsub[f]{DMC/allDMC}. Three different regimes have been identified from these results. 
Initially, the system reacts weakly to the applied fields up to $H_{c1}~=~0.15$~T (regime I). As the applied field is increased above $H_{c1}$, the correlation length $\xi_c$ along the chains decreases, reaching a minimum at $H\approx0.25$~T, and for larger fields the tendency is reversed with an increasing $\xi_c$ (regime II). 
In this regime, the peak position $l$ changes continuously from $\frac{1}{2}$ to $\frac{1}{3}$. Above $H_{c1}^*~=~0.45$~T, $l$ is fixed at $\frac{1}{3}$ and the correlation lengths change very slowly with field (regime III). The three regimes are also observed in the correlation length $\xi_b$ transverse to the chain directions. The correlation length $\xi_b$ is weakly affected by the field in regimes I and III while it reaches a maximum in regime II. This suggests that the reduced correlation length $\xi_c$ along the chain in this regime favours correlations along the $b$-axis by interchain interactions. The interchain interactions cause a different field dependence of the peak positions $k$ and $l$. In regime I, the peak position $k$ first reduces to values smaller than $\frac{1}{2}$, in contrast with $l$. In regime II and III, it then converges towards $k~=~\frac{1}{3}$.



In the ultrasound measurements for increasing fields, the transition at $H_{c1}$ separates into three different transitions below $T\approx0.3$~K, which are at $H~=~0.27$, 0.44 and 0.91~T at $T~=~100$~mK.\cite{Bidaud2016} The first transition observed at $H~=~0.27$~T corresponds to the observed minimum of $\xi_c$ in regime II. The second transition separates regimes II and III when the wavevector along the chain reaches a fixed value. A third transition is observed in ultrasound measurements at 0.91~T but no significant changes are apparent in our neutron scattering data around this field value and the nature of this transition is unknown. 

These multiple regimes are only observed at very low temperatures in ultrasound measurements. In our neutron scattering results, this is evidenced indirectly in the field dependence of the scattering intensity at fixed $\textbf{Q}~=~(0,0.66,0.67)$ shown on Fig.~\refsub[g]{DMC/allDMC}. At $T~=~70$~mK there is a sudden increase of intensity at $H_{c1}^*~=~0.45$~T, consistent with the propagation vector approaching closely $\textbf{k}_{1/3}$. In contrast, the magnetic scattering intensity at $T~=~0.2$~K and above appears at $H\approx0.3$~T and increase smoothly with field. This increase occurs close to the lower phase boundary measured in FC conditions at these temperatures and suggests a single transition.


\subsection{Partially ferromagnetic phase}

\begin{figure}[!htb]
\includegraphics[scale=1.00]{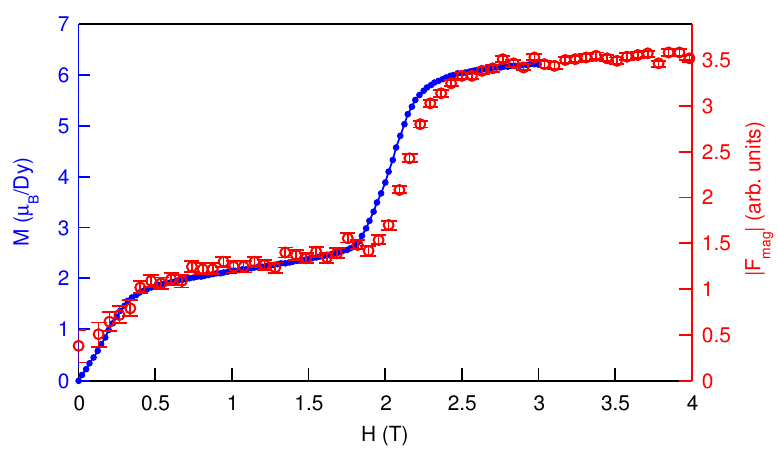}
\vspace{-0.4cm}
\caption{Left axis: Magnetization curve for $\textbf{H} \parallel \textbf{b}$ at $T~=~0.66(4)$~K showing a magnetization plateau at 1/3 of the saturation value. The data are not corrected for demagnetization effects. Right axis: Measured magnetic structure factor of the $\textbf{Q}~=~(4,0,0)$ peak for $\textbf{H} \parallel \textbf{b}$ at $T~=~0.50$~K. The small discrepancy in $H_{c2}$ between the two datasets originates from different demagnetization effects expected for samples of different shapes.}
\label{MHcurve}
\end{figure}

At magnetic fields greater than the saturation field $H_s$, a partially ferromagnetic state is reached, in which all moments on site 2 are polarized along the field direction while the moments on site 1 only have a small component along the field. Magnetization for field applied along the $b$-axis at $T~=~0.66$~K exhibit the expected plateau at $\frac{1}{3}M_s$ and a second plateau at $M_s$ corresponding to the partially ferromagnetic phase, as shown on Fig.~\ref{MHcurve}. The critical fields $H_{c1}~=~0.14(2)$~T and $H_{c2}~=~1.92(2)$~T of the \uud\ phase were extracted from the maxima of $\partial M/ \partial H$ after correction for the demagnetization effect. The saturation field $H_s~=~2.02(2)$~T at $T~=~0.66$~K was obtained from the onset of the magnetization plateau $M_s$, also after correction for the demagnetization effect. The plateaux at $\frac{1}{3}M_s$ and $M_s$ are consistent with the neutron scattering intensity at $\textbf{Q}~=~(4,0,0)$ that is sensitive to magnetization.
The phase boundary of the partially ferromagnetic order is apparent from the significant increase of the scattering intensity at $\textbf{Q}~=~(4,0,0)$ just below $H_s$ as shown on Fig.~\refsub[a]{TriCS:Peaks011_400_Hdep}.

\begin{figure}[!htb]
\includegraphics{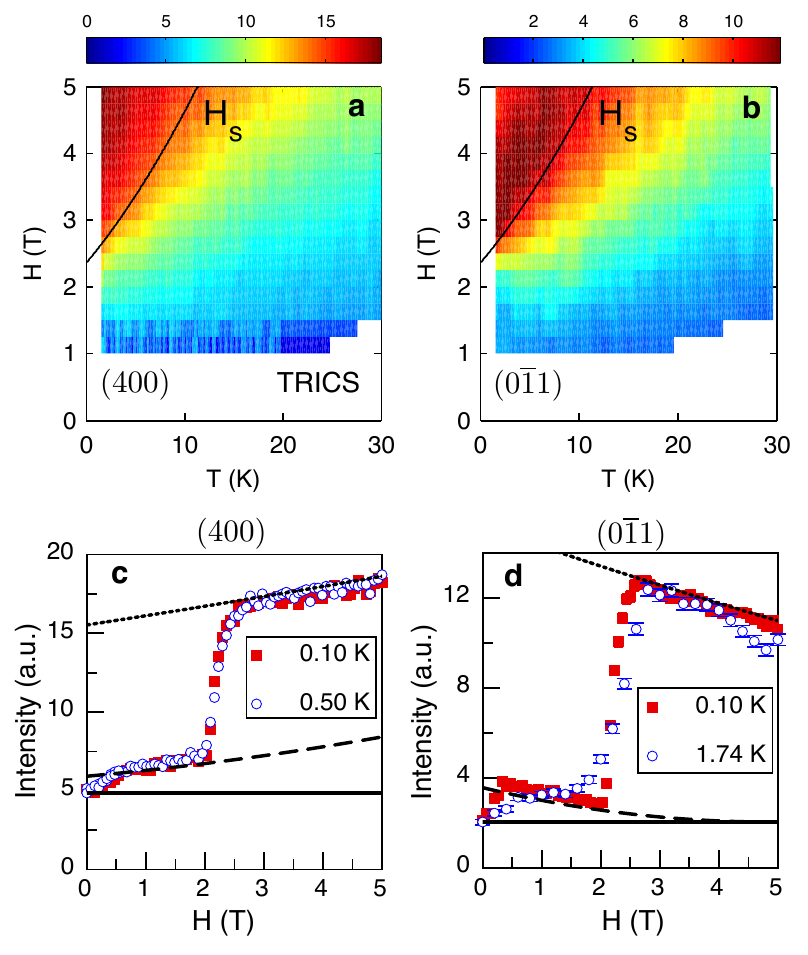}
\vspace{-0.5cm}
\caption{Phase diagram of the ferromagnetic phase as seen from the intensity of the {(a)} $\textbf{Q}~=~(4,0,0)$ and  {(b)} $\textbf{Q}~=~(0,\overline{1},1)$ peaks. Low temperature field-dependence of the intensity of the {(c)} $\textbf{Q}~=~(4,0,0)$ and {(d)} $\textbf{Q}~=~(0,\overline{1},1)$ peaks. 
The full, dashed and dotted lines are calculations for $|F_\text{nuc}|^2+|F_\text{mag}|^2$ where $F_\text{mag}~=~0$, $F_\text{mag}~=~F_\text{mag}^\text{plateau}$ and $F_\text{mag}~=~F_\text{mag}^\text{ferro}$, respectively (see text).}
\label{TriCS:Peaks011_400_Hdep}
\end{figure}

Interestingly, the scattering intensity at $\textbf{Q}~=~(0,\overline{1},1)$ is decreasing in the plateau region and the partially ferromagnetic state [Fig.~\refsub[d]{TriCS:Peaks011_400_Hdep}]. This behaviour can be explained by a field-independent moment on site 2 and a small field-dependent contribution of the moments on site 1 along the $b$-axis. The field-dependent magnetic structure factor at $\textbf{Q}~=~(0,\overline{1},1)$ and $(4,0,0)$ was calculated for this scenario. It was assumed that site 1 has a linear susceptibility in the plateau region and in the partially ferromagnetic state. It was also assumed that the order in the field induced phase is the one determined previously, in addition to a weak component along the $b$-axis on site 1 from the mentioned susceptibility.  For the partially ferromagnetic phase, it was assumed that the site 2 is fully polarized along the $b$-axis and that the moment on site 1 is $M~=~(0,1.1\mu_\text{B} +H\chi^\text{ferro}_1,0)$. A global scaling factor, a constant background and the $\chi_1$ values were adjusted to obtain a reasonable description of the data. The susceptibilities were determined to be $\chi^\text{plateau}_1~=~0.50 \mu_\text{B}/\text{T}$ and $\chi^\text{ferro}_1~=~0.25 \mu_\text{B}/\text{T}$. The calculated structure factors for the plateau and the partially ferromagnetic state are illustrated on Fig.~\refsub[c]{TriCS:Peaks011_400_Hdep}-\refsub[d]{TriCS:Peaks011_400_Hdep} by dashed and dotted lines respectively and are in good agreement with measurements.

\section{Discussion}
\subsection{Intra and interchain interactions}
The compound \sdo\ in zero magnetic field is well described in first approximation by the 1D ANNNI model and the comparison of the diffuse scattering with the model gives estimated values of $J_1\approx0.3$~meV and $J_2\approx0.2$~meV on the zig-zag chains.\cite{Gauthier2017} The 1D ANNNI model predicts the presence of the \uud\ state for longitudinal fields with antiferromagnetic $J_1$ and $J_2$.\cite{Oguchi1965,Morita1972,Rujan1983} On the other hand, transverse fields should lead to quantum fluctuations and the \uud\ state is not expected in that case.\cite{Chakrabarti1996} For fields applied along the $b$-axis in \sdo, the moments on site 1 experience a transverse field while the moments on site 2 experience a longitudinal field. Interestingly, the magnetic structure of the field-induced phase determined in section~\ref{magstruc} exhibit the \uud\ state on both inequivalent sites. The ordering on site 1 is unexpected and most likely caused by interchain interactions. Quantum fluctuations due to the transverse field should destabilize the order on site 1 but its presence suggests that, either the weak $XY$ anisotropy of site~1 weakens this effect, or the applied field is not sufficient to reach the vicinity of the quantum critical point. However, the increase of quantum fluctuations appears to be sufficient to reduce to strength of the trapping mechanism, leading to the absence of long range order in zero field,\cite{Gauthier2017} and help attain a long range field-induced phase in FC conditions.  


As discussed in our previous work,\cite{Gauthier2017} the most likely interchain interactions are the dipolar interactions and the interaction $J_b$, represented on Fig.~\refsub[b]{DMC/MagStruct}, would be the dominant one. In the field-induced magnetic structure, five out of six $J_b$ bonds minimize the interaction energy. This suggests that the field induced \uud\ state on site 2 forces the same state on site 1 through the dipolar interactions. 



Assuming a negligible contribution from site 1, the intrachain interactions on site 2 can be estimated from the accurate experimental values of the critical fields, obtained from the magnetization corrected for demagnetization, and the moment size refined from neutron scattering. For a zig-zag chain having the double N\'eel ground state in zero field, the interactions are given by $J_1~=~M(H_{c2}-H_{c1})/3~=~0.28$~meV, $J_2~=~M(H_{c2}+2H_{c1})/6~=~0.18$~meV, assuming the moment $M~=~8.3 \mu_\text{B}$, corresponding to the component along the $b$-axis on site 2. These values are in very good agreement with the ones estimated from the diffuse scattering in zero field.\cite{Gauthier2017} 





\subsection{Hysteresis and low temperature regimes}

%
%

In order to understand the multiple regimes observed at low temperatures in ZFC conditions, we considered the 1D ANNNI model in longitudinal field. The Hamiltonian is: 
\begin{equation}
{H} = \sum_i  J_1 {\hat{S}_i^z} {\hat{S}_{i+1}^z} + J_2 {\hat{S}_i^z} {\hat{S}_{i+2}^z}-h\hat{S}_i^z.
\end{equation}
This model can be solved exactly by the transfer matrix method\cite{Bhattacharyya1999} and the spin pair correlation function at large distances is given by:
\begin{equation}
G(r)=m^2 + Ae^{-\frac{r}{\xi}}\cos(qr)
\end{equation}
where $m$ is the magnetization, $A$ is a scaling factor, $\xi$ is the correlation length and $q$ is the wavenumber.\cite{Alves2000} The exact solution of this model provides analytical functions for the field-dependence of $\xi$ and $q$. These are shown for $J_1~=~0.28$~meV, $J_2~=~0.18$~meV and $M~=~8.3 \mu_\text{B}$ at various temperatures on Fig.~\refsub[a]{ANNNIField}-\refsub[b]{ANNNIField}. The wavenumber changes smoothly from its zero field value ($q\approx 0.5$) to its field-induced value ($q\approx \frac{2}{3}$). The correlation length has a minimum at an intermediate value of $q$. The field range of these changes becomes narrower with decreasing temperature. 

\begin{figure}[!htb]
\includegraphics[scale=0.7]{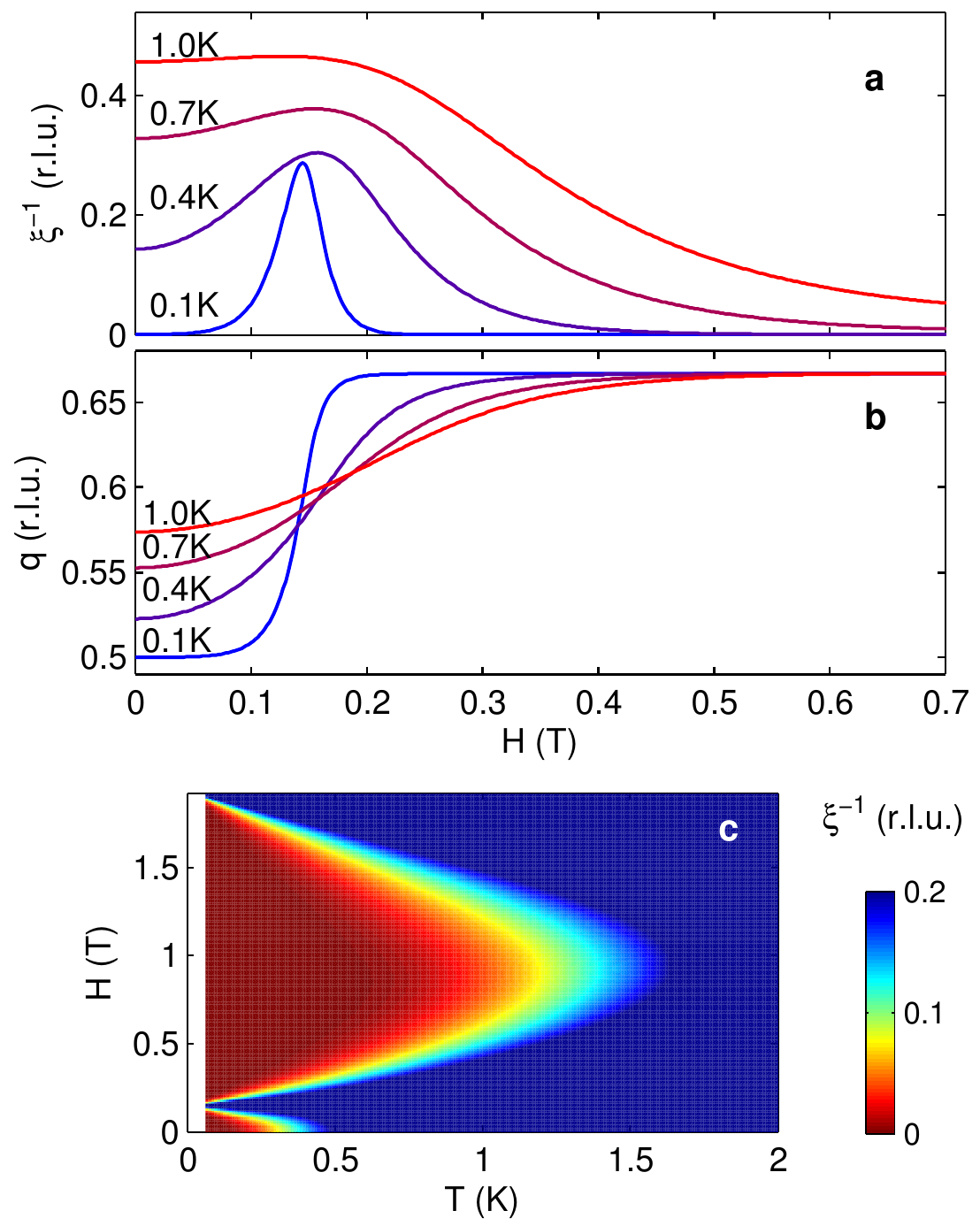}
\vspace{-0.2cm}
\caption{Field dependence of the (a) inverse correlation length $\xi^{-1}$ and (b) wavenumber $q$ according to the 1D ANNNI model in longitudinal field for various temperatures. The curves are represented in reciprocal lattice units of the zig-zag chain for $J_1~=~0.28$~meV, $J_2~=~0.18$~meV. (c) Field and temperature dependence of the inverse correlation length $\xi^{-1}$ for the same parameters.}
\label{ANNNIField}
\end{figure}

At sufficiently low temperatures the correlations are almost unaffected by the field up to a threshold value. The regime I in \sdo\ corresponds to this field range where the system is nearly field-independent. The regime II occurs when the correlation length along the chain decreases. The reduction of these correlations yields an increase of the interchain correlations, as observed by an increase in the correlation length along the $b$-axis. In the 1D ANNNI model, the decrease and increase of the correlation length at $T~=~60$~mK happens in a narrower field range than what is experimentally observed in \sdo\ (regime II). This may be due to the interchain correlations which compete with the intrachain correlations. The regime III is attained when there is no more variation in $q$ and the correlation length along the chain. The presence of multiple regimes at low temperatures are therefore understood in a simple manner with the 1D ANNNI model in longitudinal field. The theoretical temperature and field dependence of the correlation length shown in Fig.~\refsub[c]{ANNNIField} also reflects the experimental phase diagram shown in Fig.~\refsub[a]{TriCS/allData_log}.



We have previously described the absence of long range magnetic order of \sdo\ in zero field by the presence of slowly decaying defects that are trapped by a dimensionality crossover around $T^*~=~0.7$~K.\cite{Gauthier2017} The hysteretic effects measured in field appear below this characteristic temperature. This can be understood by the very slow dynamics below $T^*$ that precludes the rearrangement of large number of moments. In fact, the transitions between different orders of the ANNNI model in longitudinal fields are first order and do not lead to critical fluctuations in the proximity of the transitions. The modification of the order must therefore occur through local rearrangements of the moments and the lack of thermal fluctuations at low temperatures justifies the presence of strong hysteretic effects. The quantum fluctuations on site 1 due to the transverse field appear to be insufficient to overcome the slow thermally activated dynamics at these low temperatures. 

\section{Summary}
We presented the magnetic structure of the field-induced phase of \sdo\ for $\textbf{H} \parallel \textbf{b}$, which is consistent with the \uud\ state expected for the 1D ANNNI model in longitudinal fields. Our results show that this field-induced phase features short or long range order depending on the temperature and magnetic field history. At $T~=~60$~mK in ZFC conditions, the magnetic correlations exhibit multiple correlated regimes between the zero field short range order characterized by $\textbf{k}_{1/2}$ and the field-induced state characterized by $\textbf{k}_{1/3}$. These regimes can be qualitatively described by the 1D ANNNI model in longitudinal field. The state for fields $H>H_s$ is described as a partially ferromagnetic state with one fully polarized site and one weakly field-dependent site due to its transverse moment anisotropy. Even though our results indicate that the \sdo\ compound exhibits the main features expected by the 1D ANNNI model, the interchain couplings are the ones responsible for the system complexity leading to unconventional effects such as the transverse ferromagnetic magnetization in the field-induced phase with $\textbf{k}_{1/3}$ presented in section~\ref{magstruc}. This transverse ferromagnetism is understood from interchain interactions forcing the \uud\ state on the inequivalent site for which the moment anisotropy is transverse to the field. 

\section*{Acknowledgements}
The authors are thankful to M. Sigrist and D.L. Quintero-Castro for fruitful discussions; M. Bartkowiak and M. Zolliker for the assistance with the dilution refrigerator experiments at SINQ. 
The magnetic structure figure was generated with the SpinW package for Matlab.\cite{Toth2015}
This research received support from the Swiss National Foundation (SNF Grant No. 138018), the Natural Sciences and Engineering Research Council of Canada (Canada), the Fonds Qu\'eb\'ecois sur la Nature et les Technologies (Qu\'ebec) and the Canada Research Chair Foundation (Canada). 

%


\end{document}